\documentstyle[epsf,psfrag]{article}

\newcommand{\bda}{\begin{\displaymath}\begin{array}{rl}}
\newcommand{\eda}{\end{array}\end{displaymath}}
\newcommand{\be}{\begin{equation}}
\newcommand{\ee}{\end{equation}}
\newcommand{\bdm}{\begin{displaymath}}
\newcommand{\edm}{\end{displaymath}}
\newcommand{\bea}{\begin{eqnarray}}
\newcommand{\eea}{\end{eqnarray}}

\newcommand{\fs}{\; .}
\newcommand{\co}{\; ,}

\newcommand{\lvac}{\langle 0|\,}
\newcommand{\rvac}{\,|0\rangle}

\newcommand{\ubar}{\overline{\rule[0.42em]{0.4em}{0em}}\hspace{-0.5em}u}
\newcommand{\dbar}{\,\overline{\rule[0.65em]{0.4em}{0em}}\hspace{-0.6em}d}

\newcommand{\Mbar}{\hspace{0.1cm}\overline{\rule[0.7em]{0.8em}{0em}}\hspace{-1em}M}
\newcommand{\ChPT}{$\chi$PT }
\newcommand{\lbar}{\bar{\ell}}
\newcommand{\Kbar}{\,\overline{\rule[0.62em]{0.5em}{0em}}\hspace{-0.8em}K}

\begin{document}

\begin{center}{\LARGE\bf On the low energy end of the \\ \rule{0cm}{0.6cm}QCD spectrum}
 
 \vspace{1cm}
H. Leutwyler\\ Institute for
  Theoretical Physics, University of Bern\\ Sidlerstr.\ 5, CH-3012 Bern,
  Switzerland
\end{center}

\begin{abstract}
The experimental results on the $K_{e4}$ and $K_{3\pi}$ decays, those on pionic atoms and recent work on the lattice confirm the predictions obtained on the basis of $\chi$PT. As a result, the energy gap of QCD is now understood very well and there is no doubt that the expansion in powers of the two lightest quark masses does represent a very useful tool for the analysis of the low energy structure. Concerning the expansion in powers of $m_s$, however, the current situation leaves much to be desired. While some of the lattice results indicate, for instance, that the violations of the Okubo-Iizuka-Zweig rule in the quark condensate and in the decay constants are rather modest, others point in the opposite direction. In view of the remarkable progress being made with the numerical simulation of light quarks, I am confident that the dust will settle soon, so that the effective coupling constants that govern the dependence of the various quantities of physical interest on $m_s$  can reliably be determined, to next-to-next-to-leading order of the chiral expansion. 

The range of validity of $\chi$PT can be extended by means of dispersive methods. The properties of the physical states occurring in the spectrum of QCD below $K\Kbar$ threshold can reliably be investigated on this basis. In particular, as shown only rather recently, general principles of quantum field theory lead to an exact formula that expresses the mass and width of resonances in terms of obser\-vable quantities. The formula removes the ambiguities inherent in the analytic continuation from the real axis into the complex plane, which plagued previous determinations of the pole positions of broad resonances. The application to the $\pi\pi$ partial wave amplitude with $I=\ell=0$ shows that there is a resonance in this channel, at $M_\sigma-\frac{i}{2}\Gamma_\sigma\simeq441-i\,272$ MeV: the lowest resonance of QCD carries the quantum numbers of the vacuum. 
\begin{center}
{\it Talks given at QCD08, Montpellier, France, July 2008 and\\ Confinement8, Mainz, Germany, September 2008}
\end{center}
\end{abstract}

\protect{\newpage}

 \begin{center}
 \mbox{ \it
In memoriam Jan Stern, 29. 6. 1942 -- 2. 7. 2008}

\vspace{0.5cm}
 \mbox{\epsfysize=7.4cm \epsfbox{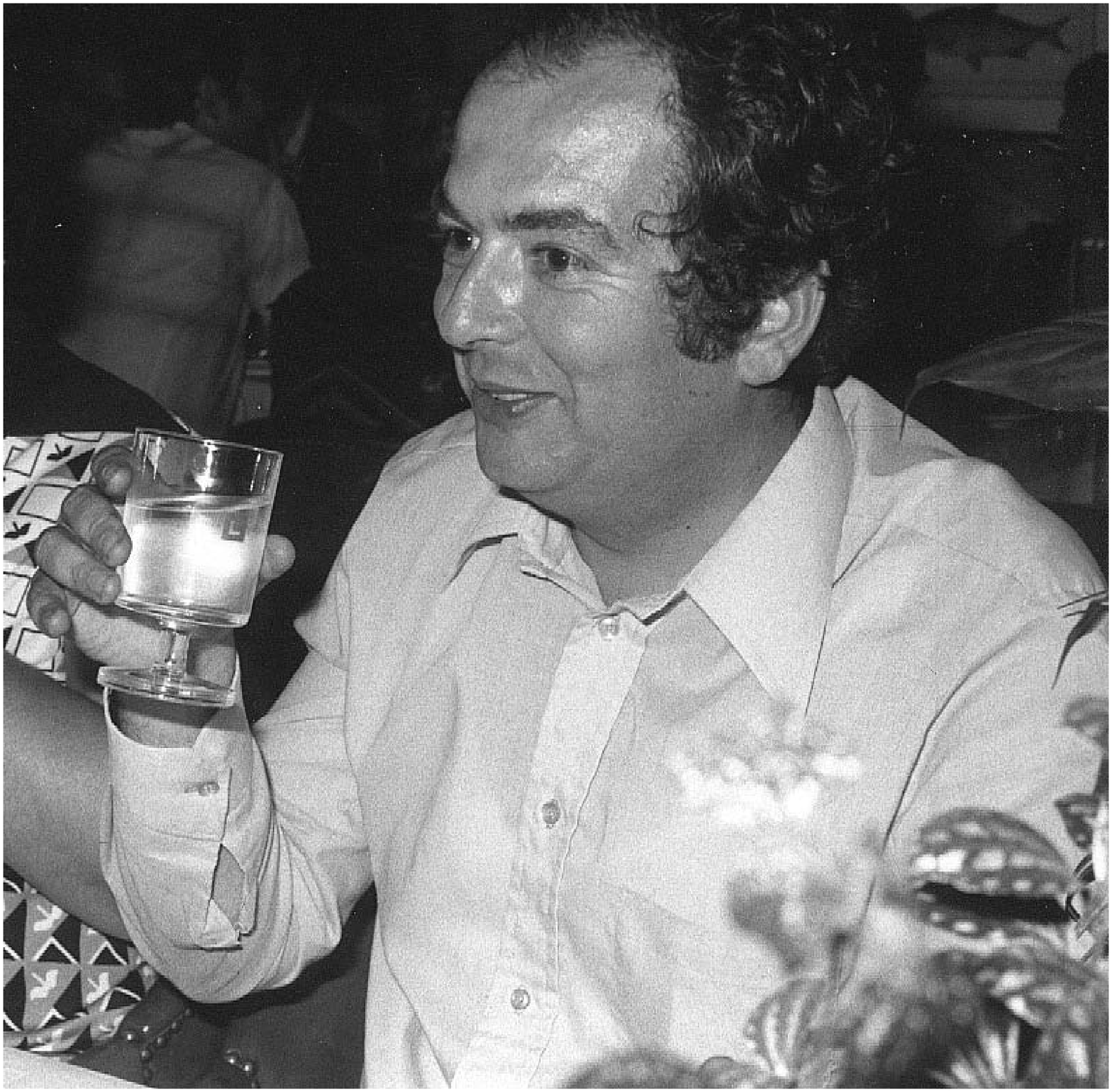}}
  
\vspace{1cm}  
\end{center}
 
I dedicate this report to the memory of Jan Stern. He was born in Prague, in 1942. When the Soviet tanks crushed his hopes for a continuation of the development initiated in the spring of 1968, he left his home country and joined the Theory Division of the Institut de Physique Nucl\'eaire at Orsay. I had the privilege of intensely collaborating with him over many years. Jan was very gifted and he wanted to know. He pursued the problems encountered in the course of his research with contagious enthusiasm -- it was a true pleasure to discuss physics with him, sharing insights, critically examining new ideas, sharpening questions, disputing different points of view, writing papers, \ldots 

Unfortunately, a cancer started affecting his life a few years ago. He participated in the symposium which his colleagues organized at Orsay in his honour, at the beginning of June. My impression was that he enjoyed the event and took pleasure chatting with his friends. A month later, however,  he passed away -- a sad fact and a great loss for our community.   

\newpage

\section{Introduction}\label{sec:intro}
In my opinion, QCD is the most satisfactory part of the Standard Model. In fact, in the limit where the quark masses are set equal to zero, QCD is the ideal of a theory: it does not contain a single dimensionless free parameter. Since QCD is asymptotically free, the coupling between the quarks and gluons becomes strong at low energies, so that  the perturbative expansion in powers of the coupling constant fails. Accordingly, the progress made in understanding the low energy properties of QCD has
been very slow. A large fraction of the papers written in this field does not
concern QCD as such, but models that resemble it in one way or the other:
constituent quarks, NJL-model, linear $\sigma$ model, hidden local symmetry,
AdS/CFT and many others. Some of these may be viewed as simplified versions of
QCD that do catch some of the salient features of the theory at the
semi-quantitative level, but none provides a basis for a coherent
approximation scheme that would allow us, in principle, to solve QCD.
 
In the following, I do not consider such substitutes, but discuss the progress made in understanding the low energy structure of QCD. The key words in this context are $\chi$PT, lattice methods, experiment and dispersion theory.  For recent reviews of the developments in these four domains, I refer to \cite{Ecker,Necco,Bloch,Lisbon}. A more detailed account of the topics discussed below is given in \cite{HL Erice 2007}.

At low energies, the main characteristic of QCD is that the energy gap is remarkably
small, $M_\pi\simeq $ 140 MeV. More than 10 years before the discovery of QCD,
Nambu \cite{Nambu} found out why that is so: the gap is small because the
strong interactions have an approximate chiral symmetry. Indeed, QCD does have
this property: for yet unknown reasons, two of the quarks happen to
be very light. The symmetry is not perfect, but nearly so: $m_u$ and $m_d$ are
tiny. The mass gap is small because the symmetry is ``hidden'' or
``spontaneously broken'': for dynamical reasons, the ground state of the
theory is not invariant under chiral rotations, not even approximately. The
spontaneous breakdown of an exact Lie group symmetry gives rise to strictly
massless particles, ``Goldstone bosons''. In QCD, the pions play this role:
they would be strictly massless if $m_u$ and $m_d$ were zero, because the
symmetry would then be exact. The only term in the Lagrangian of QCD that is not
invariant under the group SU(2)$\times$SU(2) of chiral rotations
is the mass term of the two lightest quarks, $m_u\,\ubar u+m_d \,\dbar
d$. This term equips the pions with a mass. Although the
theoretical understanding of the ground state is still poor, we do have very
strong indirect evidence that Nambu's conjecture is right -- we know why the
energy gap of QCD is small.

\section{Lattice results for $M_\pi$ and $F_\pi$}\label{sec:lat}
As pointed out by Gell-Mann, Oakes and Renner \cite{GMOR}, the square of the
pion mass is proportional to the strength of the symmetry breaking,
\bdm M_\pi^2\propto (m_u+m_d)\,.\edm   This property can now be checked on the lattice, where -- in principle -- the quark masses can be varied at will. In
view of the fact that in these calculations, the quarks are treated
dynamically, the quality of the data is impressive. The masses are
sufficiently light for \ChPT to allow a meaningful extrapolation to the
quark mass values of physical interest. The results indicate that the ratio
$M_\pi^2/(m_u+m_d)$ is nearly constant out to values of $m_u, m_d$ that are
about an order of magnitude larger than in nature. According to Gell-Mann, Oakes and Renner, this ratio is related to the quark condensate. The Banks-Casher relation, which connects the quark condensate with the spectral density of the Dirac operator at small eigenvalues \cite{Banks & Casher}, is now also accessible to a numerical evaluation on the lattice \cite{Giusti Mainz}.

The Gell-Mann-Oakes-Renner relation corresponds to the leading term in the
expansion in powers of the quark masses. At next-to-leading order, the
expansion in powers of $m_u,m_d$ (mass of the strange quark kept fixed at the physical value) contains a logarithm: 
\be\label{eq:Mpi one loop} M_\pi^2=
M^2\left\{1 +\!\frac{M^2}{32\pi^2 F_\pi^2}\, \ln
  \frac{M^2}{\Lambda_3^2}\!+\!O(M^4)\right\}\co\ee
where $M^2\equiv B(m_u+m_d)$ stands for the term linear in the quark masses. 
Chiral symmetry fixes the coefficient of the logarithm in
terms of the pion decay constant $F_\pi\simeq 92.4$ MeV, but does not determine the scale $\Lambda_3$. An estimate for this scale was obtained more than 20 years
ago \cite{GL SU2}, on the basis of the SU(3) mass formulae for the pseudoscalar
octet: $\lbar_3\equiv \ln \Lambda_3^2/M_\pi^2= 2.9\pm 2.4$. Several collaborations have now managed to determine the scale $\Lambda_3$ on the lattice -- for an overview, I refer to \cite{Necco}. The result of the
RBC /UKQCD collaboration, $\lbar_3=3.13 \pm 0.33_{\,\mbox{\tiny stat}}\pm 0.24_{\,\mbox{\tiny syst}}$ \cite{RBC/UKQCD}, for instance, which concerns 2+1 flavours and includes an estimate of the systematic errors, is perfectly consistent with, but
considerably more accurate than our old estimate. 

The expansion of $F_\pi$ in powers of $m_u,m_d$ also contains a logarithm at NLO.  The coupling constant relevant in that case is denoted by $\lbar_4$. A couple of years ago, we obtained a rather accurate result for this quantity, from a dispersive analysis of the scalar form factor: $\lbar_4=4.4\pm0.4$  \cite{CGL} (for details,  I refer to \cite{ACCGL}). The lattice determinations of $\lbar_4$ have reached comparable accuracy and are consistent with the dispersive result \cite{Necco}. The scalar form factor is now also accessible to an evaluation on the lattice \cite{Kaneko}.  

\section{Precision experiments at low energy}\label{sec:Precision experiments}
The hidden symmetry not only controls the size of the energy gap, but also
determines the interaction of the Goldstone bosons at low energies, among
themselves, as well as with other hadrons. In particular, as pointed out by
Weinberg \cite{Weinberg 1966}, the leading term in the chiral expansion of the
S-wave $\pi\pi$ scattering lengths (tree level of the effective theory) is
determined by the pion decay constant. In the meantime, the chiral perturbation series of the scattering amplitude has been worked out to NNLO \cite{BCEGS} and, matching the chiral and dispersive representations, a very sharp prediction for the scattering lengths was obtained: $a_0^0=0.220(5)$, $a_0^2=-0.0444(10)$ \cite{CGL}. 

While the $K_{e4}$ data of E865 \cite{E865}, the DIRAC experiment \cite{DIRAC} and the NA48 data on the cusp in $K\rightarrow 3\pi$ \cite{NA48 cusp} all confirmed the theoretical expectations, the most precise source of information, the beautiful $K_{e4}$ data of NA48 \cite{NA48 paper on Ke4}, gave rise to a puzzle. This experiment exploits the fact that -- if the electromagnetic interaction and the difference between $m_u$
and $m_d$ are neglected -- the relative phase of the form factors describing
the decay $K\rightarrow e\nu\pi\pi$ coincides with the difference
$\delta_0^0-\delta_1^1$ of scattering phase shifts (Watson theorem). At the precision achieved, the data on the form factor phase did not agree with the theoretical prediction
for the phase shifts. 
\begin{figure}[thb]\centering
  \mbox{\epsfysize=8cm \hspace{-0.5cm}\rotatebox{-90}{\epsfbox{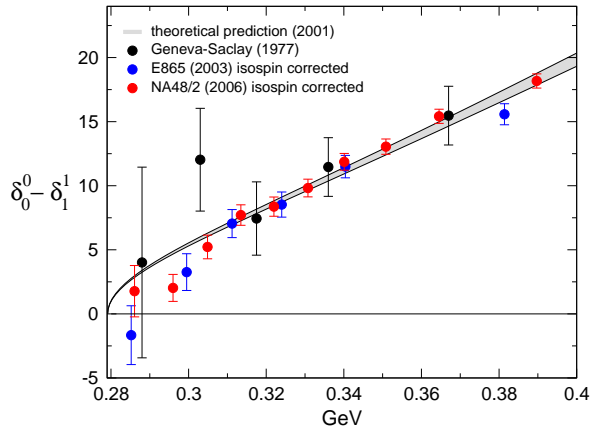}}}
\caption{\label{fig:delta0011}Comparison of $K_{e4}$ data with the prediction for
  $\delta^0_0-\delta^1_1$}  
\end{figure}

The origin of the discrepancy was identified by Colangelo, Gasser and Rusetsky
\cite{Colangelo Gasser Rusetsky}. The problem has to do with the fact that
a $K^+$ may first decay into $e^+\nu\hspace{0.5mm}\pi^0\pi^0$ -- the pair of neutral pions then undergoes scattering and winds up as a charged pair. The mass difference
between the charged and neutral pions affects this process in a pronounced
manner: it pushes the form factor phase up by about half a degree -- an
isospin breaking effect, due almost exclusively to the electromagnetic
interaction. 

Figure \ref{fig:delta0011} shows that the discrepancy disappears if the NA48 data
on the relative phase of the form factors are corrected for isospin breaking. In view of the fact that the prediction does not contain any free parameters, the agreement amounts to a strong test of the theory. 

The low energy theorem for the scalar radius of the pion correlates the two S-wave scattering lengths to a narrow strip \cite{CGL}. If this correlation is used, the $K_{e4}$ data determine $a^0_0$ to the same precision as the theoretical prediction and hit it on the nail: $a_0^0=0.220(5)(2)$ \cite{Bloch}. The analysis of the same data by Yndur\'ain et al.\  \cite{Pelaez et al Lisbon scattering lengths}, which does not rely on the low energy theorem for the scalar radius, but instead uses phase shifts extracted from $\pi N$ reactions, confirms this result within errors. Moreover, their analysis leads to a phenomenological determination of the exotic scattering length $a_0^2$, which is in excellent agreement with the theoretical prediction. I conclude that the puzzle is gone: the experimental results on $K_{e4}$ decay confirm the theory to remarkable precision.

The corrections to Weinberg's low energy theorem for $a_0^0, a_0^2$ are dominated by the effective coupling constants $\bar{\ell}_3,\bar{\ell}_4$ discussed above -- if these are known, the scattering lengths can be calculated within small uncertainties \cite{CGL}. The lattice results for these couplings can therefore be translated into corresponding results for $a_0^0,a_0^2$. For a recent review of the values for the scattering lengths obtained in this way, I refer to \cite{Ecker}. Since the lattice results confirm the values of $\lbar_3,\lbar_4$ used in \cite{CGL}, it does not come as a surprise that the results for the scattering lengths also agree with our predictions, but some of these are considerably more precise. NPLQCD, for instance, quotes the outcome for the exotic scattering length $a_0^2$ to an accuracy of 1\% \cite{NPLQCD PR 2008}, systematic errors included. The result is obtained by analyzing mixed-action data by means of $\chi$PT to NLO. 
 \section{\boldmath Expansion in powers of $m_s$}\label{sec:ms}
The examples discussed above all concern the effective theory based on SU(2)$\times$SU(2), where the quantities of interest are expanded in powers of $m_u,m_d$, while $m_s$ is kept fixed at the physical value. The corresponding effective coupling constants $F,B,\ell_1,\ldots$ are independent of $m_u$ and $m_d$, but do depend on $m_s$. 
Chiral symmetry determines the quark condensate in terms of  $F,B$:  
\be \Sigma \equiv \lvac \ubar u\rvac|\rule[-2mm]{0mm}{5mm}_{\;m_u,m_d\rightarrow 0}=F^2B\fs\ee 
The expansion of these quantities in powers of $m_s$ can be worked out in the framework of the effective theory based on SU(3)$\times$SU(3).  For $F$ and $\Sigma$, the expansion starts with\footnote{The contributions of $O(m_s^2)$ are also known explicitly, not only for $F,B,\Sigma$, but also for the coupling constants $\ell_1,\ldots$ which specify the effective Lagrangian at NLO  \cite{GHIS,Kaiser Mpi}.}
 \bea\label{eq:FSigma} F\hspace{-0.1cm}&\hspace{-0.1cm}=\hspace{-0.1cm}&\hspace{-0.1cm}F_0\left\{ 1+\frac{8\Mbar^2_K}{F_0^2}L_4^r-\bar{\mu}_K+O(m_s^2)\right\}\,,\\
 \Sigma\hspace{-0.1cm}&\hspace{-0.1cm}=\hspace{-0.1cm}&\hspace{-0.1cm} \Sigma_0\left\{1+\frac{32\Mbar^2_K}{F_0^2}L_6^r-2\bar{\mu}_{K}-\bar{\mu}_{\eta}+O(m_s^2)\right\} \,.\nonumber\eea
The constants $F_0,\Sigma_0$ represent the values of $F,\Sigma$ in the limit $m_s\rightarrow 0$. At NLO, only the coupling constants $L_4,L_6$ of the chiral SU(3)$\times$SU(3) Lagrangian enter, weighted with the square of the kaon mass in the limit $m_u,m_d=0$, which I denote by $\Mbar_K$. In this limit, the octet of Goldstone bosons contains only three different mass values: $\Mbar_\pi=0$,  $\Mbar_K$ and $\Mbar_\eta$. To the accuracy relevant in the above formulae, we have $\Mbar_K^2=B_0\,m_s$, $\Mbar_\eta^2=\frac{4}{3}B_0\,m_s$. Up to corrections of higher order, $\Mbar_K$ and $\Mbar_\eta$ may be expressed in terms of the physical masses as
\be \Mbar_K^2=M_K^2-\mbox{$\frac{1}{2}$}M_\pi^2\,,\hspace{0.5cm}
\Mbar_\eta^2=\mbox{$\frac{4}{3}$}M_K^2-\mbox{$\frac{2}{3}$}M_\pi^2\fs\ee
The chiral logarithms occurring in the above formulae may be expressed in terms of the function
\be \bar{\mu}_P=\frac{\Mbar_P^2}{32 \pi^2 F_0^2}\ln\frac{\Mbar_P^2}{\mu^2}\,,\hspace{0.5cm}P=K,\eta\,.\ee
They involve the running scale $\mu$ at which the chiral perturbation series is renormalized, but the scale dependence of the renormalized coupling constants $L_4^r,L_6^r$ ensures that the expressions in the curly brackets of equation (\ref{eq:FSigma}) are scale independent.

The coupling constants $L_4, L_6$ as well as the loop graphs responsible for the chiral logarithms represent effects that violate the Okubo-Zweig-Iizuka rule. In the large $N_c$ limit, the quantities $F,B,\Sigma$ become independent of $m_s$, so that the ratios $F/F_0, B/B_0, \Sigma/\Sigma_0$ tend to 1. If the OZI rule is a good guide in the present context, then these ratios should not differ much from 1. For a discussion of the implications of large OZI violations in these ratios, see \cite{Descotes lattice07}. The paramagnetic inequalities of Stern et al. \cite{paramagnetic} indicate that the sign of the deviations $F/F_0-1$ and $\Sigma/\Sigma_0-1$ is positive.

\section{Violations of the OZI rule ?}\label{sec:OZI}
\begin{figure}[thb] \centering \mbox{\epsfysize=8.4cm \rotatebox{-90}{\epsfbox
{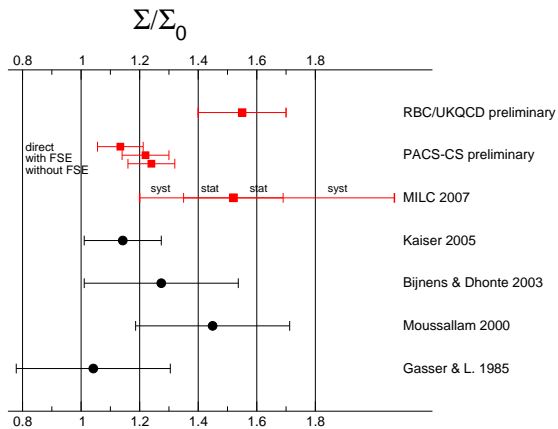} }}
\caption{\label{fig:Sigma}Violation of the OZI rule in the quark condensate}
\end{figure}

Figure \ref{fig:Sigma} compares recent lattice results for the dependence of the condensate on $m_s$ \cite{MILC,PACS,RBC/UKQCD Lattice 07} with phenomenological estimates found in the literature \cite{GL SU3,Moussallam 2000,Bijnens & Dhonte 2003,Kaiser Trento & Montpellier} (the latter are calculated from the values quoted for the running coupling constant $L_6$, using the relation (\ref{eq:FSigma}) with $F_0=F_\pi$). The errors shown exclusively account for the quoted uncertainties in the coupling constants, while those arising from the corrections of $O(m_s^2)$ are neglected. For details, I refer to  \cite{HL Erice 2007}. The plot shows that the uncertainties in the phenomenological estimates are large. Unfortunately, the lattice results are not yet conclusive, either. Note that some of these are preliminary and do not include an estimate of the systematic errors. 

In contrast to the lattice results for the condensate, those for the constant $B$ are consistent with one another. The values obtained for $B/B_0$ do not indicate a large violation of the OZI rule. This implies that the discrepancy seen in the lattice results for $\Sigma=F^2B$ originates in the factor $F^2$. Indeed, some of the values quoted for $F/F_0$  are puzzling, for the following reason. The quantity $F_\pi$ represents the pion wave function at the origin. The value of $F_K$ is somewhat larger, because one of the two valence quarks is heavier than in the case of the pion. Hence it moves more slowly, so that the wave function is more narrow and thus higher at the origin:  
$F_K/F_\pi= 1.192(7)\; \cite{Bernard & Passemar}$. 

If the value of $F/F_0$ was larger than this, we would have to conclude that the wave function is more sensitive to the mass of the sea quarks than to the mass of the valence quarks. I do not see a way to rule this logical possibility out, but it is counter intuitive and hence puzzling. For the time being, the only conclusion to draw is that the lattice results confirm the paramagnetic inequalities and indicate that the constant $B$ -- the leading term in the expansion of $M_\pi^2$ in powers of $m_u$ and $m_d$ -- does obey the OZI rule. Some of the data indicate that this rule approximately holds also for $F/F_0$, but others suggest rather juicy violations in that case. Since this is one of the hot spots in current research, I am confident that the discrepancies will soon be resolved.  

\section{Exact formula for resonances}\label{sec:Pole formula}
The preceding discussion concerns the domain where $\chi$PT provides useful information. In the remainder of this brief report, I consider an issue outside that domain: the poles of the S-matrix. Their positions represent universal properties of QCD, which are unambiguous even if the width of the resonance turns out to be large, but they are outside the reach of perturbation theory.   

The progress recently made in the determination of the lowest resonance of QCD derives from an exact formula that expresses the poles of the $\pi\pi$ S-matrix elements in terms of physical quantities \cite{CCL}. The proof of this formula relies on the well known fact that poles on the second sheet of a partial wave amplitude give rise to zeros of the S-matrix on the first sheet and vice versa. On the first sheet, the dispersive representation established by Roy \cite{Roy} represents the real part of the partial waves as a sum of integrals over the imaginary parts (the sum extends over all partial waves). \begin{figure}[thb]\centering \mbox{\epsfysize=8.4cm \rotatebox{-90}{\epsfbox
{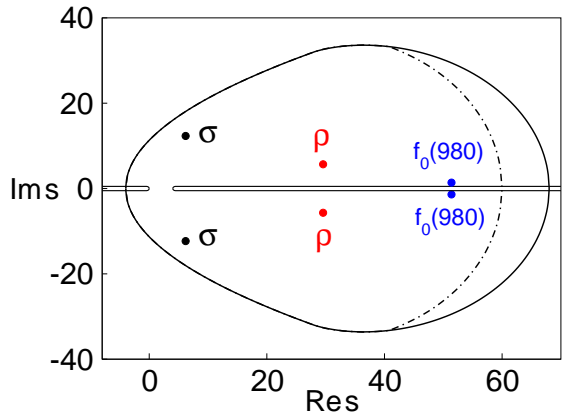} }}
\caption{\label{fig:domain}Domain of validity of the Roy equations and positions of the lowest few resonances of QCD (real and imaginary parts of $s$ in units of $M_\pi^2$).}
\end{figure}
Using known results of general quantum field theory, we have shown that these equations also hold for complex values of $s$, in the limited region of the first sheet shown in figure \ref{fig:domain} (the full line indicates the boundary of the region obtained from Mandelstam-analyticity, while the dash-dotted one exclusively relies on general principles of quantum field theory). The pole formula is an immediate consequence of these properties of the partial waves.

The values quoted by the Particle Data Group for the pole position of the lowest isoscalar spin zero resonance, the $f_0(600)$ -- commonly referred to as the $\sigma$ --  cover a very broad range. One of the reasons is that all but one of these either rely on models or on the extrapolation of simple parametrizations: the data are represented in terms of suitable functions on the real axis and the position of the pole is determined by continuing this representation into the complex plane. If the width of the resonance is small, the ambiguities inherent in the choice of the parametrization do not significantly affect the result, but the width of the $\sigma$ is not small. For a thorough discussion of the sensitivity of the pole position to the freedom inherent in the choice of the parametrization, I refer to \cite{Caprini}. Our method does not suffer from such ambiguities, because the Roy equations explicitly specify the analytic continuation in terms of physical quantities, namely the imaginary parts of the partial waves on the real axis and the two S-wave scattering lengths $a_0^0, a_0^2$, which were discussed in section \ref{sec:Precision experiments} (these enter the Roy equations as subtraction constants).

\section{The lowest resonance of QCD}\label{sec:Lowest resonance}
In \cite{CCL}, we evaluated the pole formula numerically, for resonances with the quantum numbers of the vacuum, $I=\ell=0$. For these, the element $S^0_0(s)$ of the S-matrix is relevant. Since $S^0_0(s)$ is a real-analytic function, the zeros occur in pairs of complex conjugate values. The numerical evaluation is discussed in detail in \cite{Lisbon,CCL}. For our central representation of the scattering amplitude, we find that, in the region where the Roy equations are valid, the function $S^0_0(s)$ has two pairs of zeros: one at $\sqrt{s}= 441 \pm i\, 272$ MeV, the other $\sqrt{s}=1001\pm i\,14$ MeV. While the first corresponds to the $\sigma$, the second zero represents the well-established resonance $f_0(980)$. Our analysis sheds little light on the properties of the $f_0(980)$, because the location of the zero is sensitive to the input used for the elasticity $\eta^0_0(s)$ -- the shape of the dip in $\eta^0_0(s)$ and the position of the zero represent two sides of the same coin.  For comparison, the third pair of points in figure \ref{fig:domain} indicates the position of the $\rho$, which does not drill a zero into $S_0^0(s)$, but into $S^1_1(s)$.

We are by no means the first to find a resonance in the vicinity of the above position. In the list of papers quoted by the Particle Data Group \cite{PDG 2008}, the earliest one with a pole in this ball park appeared more than 20 years ago \cite{Beveren}.  What is new is that we can perform an error calculation, because our method is free of the systematic theoretical uncertainties inherent in models and parametrizations. For a detailed discussion of the error analysis, I refer to \cite{Lisbon}. The net result for the position of the pole on the lower half of the second sheet reads \cite{CCL}:
\be\label{eqmsigma} M_\sigma -i\,\mbox{$\frac{1}{2}$}\Gamma_\sigma=441\, \rule[-0.2em]{0em}{1em}^{+16}_{-\,8}-
\,\mbox{i}\;272\,\rule[-0.2em]{0em}{1em}^{+\,9}_{-12.5}\;\mbox{MeV}\;.\ee
The error bars account for all sources of uncertainty and are an order of magnitude smaller than the range $M_\sigma -i\,\mbox{$\frac{1}{2}$}\Gamma_\sigma$ = (400 - 1200) - i (250 - 500) MeV quoted by the Particle Data Group \cite{PDG 2008}: the position of the lowest resonance of QCD can now be calculated reliably and quite accurately \cite{Descotes Moussallam kappa}.

\section*{Acknowledgments}I thank Stefan Narison, Nora Brambilla and Matthias Neubert for a very pleasant stay at Montpellier and Mainz.


\begin{thebibliography}{99}
\bibitem{Descotes Moussallam kappa}
S.~Descotes-Genon and B.~Moussallam,
  %``The K0*(800) scalar resonance from Roy-Steiner representations of pi K
  %scattering,''
  Eur.\ Phys.\ J.\  C {\bf 48} (2006) 553, have extended the analysis to
the mesonic sector of QCD with strangeness $\pm1$, with a remarkably sharp outcome for the mass and the width of the $\kappa$.
\end{thebibliography}
\end{document}